# Using profiles based on hydropathy properties to define essential regions for splicing.

Anatoly Ivashchenko[1], Galina Boldina[1], Aizhan Turmagambetova [1], Mireille Régnier[2]

1 - KazNU named after al-Farabi- The Kazakh National University named after al-Farabi, Almaty, Kazakhstan
2 – INRIA (Institut National  de Recherche en Informatique et en Automatique), BP 105, Le Chesnay, France

a_ivashchenko@mail.ru
gboldina@mail.ru
aichyck@mail.ru
mireille.regnier@inria.fr

## ABSTRACT

We define new profiles based on hydropathy properties and point out specific profiles for regions surrounding splice sites. We built a set $T$ of flanking regions of genes with 1-3 introns from 21$^{st}$ and 22$^{nd}$ chromosomes. These genes contained 313 introns and 385 exons and were extracted from _GenBank_. They were used in order to define hydropathy profiles. Most human introns, around 99.66%, are likely to be U2-type introns. They have highly degenerate sequence motifs  and many different sequences can function as U2-type splice sites. Our new profiles allow to identify regions which have conservative biochemical features that are essential for recognition by spliceosome.

We have also found differences between hydropathy profiles for U2 or U12-types of introns on sets of spice sites extracted from _SpliceRack_ database in order to distinguish GT–AG introns belonging to U2 and U12-types. Indeed, intron type cannot be simply determined by the dinucleotide termini. We show that there is a similarity of hydropathy profiles inside intron types. On the one hand, GT–AG and GC–AG introns belonging to U2-type have resembling hydropathy profiles as well as AT–AC and GT–AG introns belonging to U12-type. On the other hand, hydropathy profiles of U2 and U12-types GT–AG introns are completely different.

Finally, we define and compute a pvalue; we compare our profiles with the profiles provided by a classical method, _Pictogram_.

There are at least two classes of pre-mRNA introns, based on the splicing machineries that catalyze the reaction: U2 snRNP-dependent introns make up the majority of all introns. These introns consist of three subtypes, according to their terminal dinucleotides: GT–AG, GC–AG and AT–AC introns. U12 snRNP-dependent introns are the minor class of introns. These introns mainly consist of two subtype s, as defined by their terminal dinucleotides: AT–AC and GT–AG introns. In addition, a small fraction of the U12-type introns exhibit other terminal dinucleotides (5–7). This raises the question of how to distinguish U12-type introns from U2-type introns, because intron type can not be simply determined by the dinucleotide termini.

## INTRODUCTION

Pre-mRNA splicing is a nuclear process that is conserved across eukaryotes [Collins and Penny, 2005]. The spliceosome recognizes conserved sequences at the exon–intron boundaries, namely the 5' splice site (5'ss) and the 3' splice site (3'ss) (Fig.1). In addition, a third conserved intronic sequence that is known to be functionally important in splicing is the so-called branch point site (BPS) which is usually located very close to the end of the intron, at most 40 nucleotides before the terminal AG dinucleotide.

The splicing mechanism involves the following steps:

i.   cleavage at the 5′ splice junction;
ii.  nucleolytic attack by the terminal G nucleotide of the splice donor site at the invariant A of the branch site to form a lariat-shaped structure;
iii. cleavage at the 3′ splice junction, leading to release of the intronic RNA as a lariat, and splicing of the exonic RNA segments.

The above reactions are mediated by a large RNA-protein complex, the spliceosome, which consists of five types of snRNA (small nuclear RNA) and more than 200 proteins ( Stanley and Guthrie, 1998). Each of the snRNA molecules is attached to specific proteins to form snRNP particles and the specificity of the splicing reaction is established by RNA-RNA base-pairing between the RNA transcript and snRNA molecules. (Human Molecular Genetics. 3rd Edition. Tom Strachan, Andrew Read. 12/4/2003. 696 pages.)

There are at least two classes of pre-mRNA introns, based on the splicing machineries that catalyze the reaction. U2 snRNP-dependent introns make up the majority of all introns and are excised by spliceosomes containing the U1, U2, U4, U5 and U6 snRNPs. These introns consist of three subtypes, according to their terminal dinucleotides: GT–AG, GC–AG and AT–AC introns. U12 snRNP-dependent introns are the minor class of introns and are excised by spliceosomes containing U11, U12, U4atac, U6atac and U5 snRNPs. These introns mainly consist of two subtypes, as defined by their terminal dinucleotides: AT–AC and GT–AG introns. In addition, a small fraction of the U12-type introns exhibit other terminal dinucleotides [Wu and Krainer 1997; Levine and Durbin, 2001; Dietrich et al., 2005]. Whereas U2-type introns have been found in virtually all eukaryotes [Collins and Penny, 2005] and comprise the vast majority of the splice sites found in any organism, U12-type introns have only been identified in vertebrates, insects, jellyfish and plants [Burge et.al, 1998].

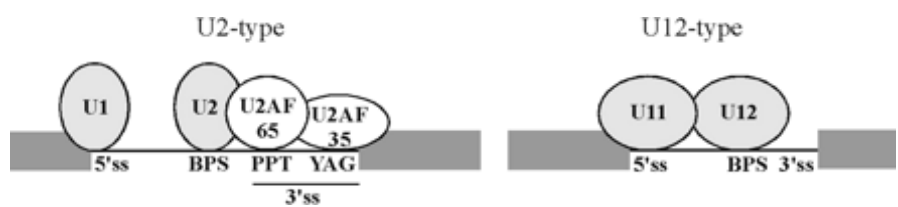

**Fig. 1. Initial steps in splice-site selection.**

The dark bars stand for exons, while the lines represent introns. ss stands for splice sites. AG is the 3' terminus of the intron, where Y is a pyrimidine. PPT is the poly-pyrimidine tract. BPS is the branch point sequence, U1, U2, U11 and U12 are snRNPs. U2AF65 and U2AF35 are protein splicing factors.

U2-type intron splicing initially involves base pairing of U1 snRNA to the 5'ss and U2 snRNA to the BPS [Brow, 2002] (Fig.1). The base pairing of U2 snRNA to the BPS is facilitated by the binding of the large subunit of the U2 Auxiliary Factor (U2AF65) to the poly-pyrimidine tract (PPT) located immediately upstream of the intron 3' terminus, and binding of the small subunit (U2AF35) to the 3' terminal AG dinucleotide of the intron [Moore, 2000; Reed, 2000]. Following the initial recognition of the splice sites by the U1 and U2 snRNPs, the U4/U6/U5 tri-snRNP is recruited to the splice site leading to the two catalytic steps of splicing [Staley and Guthrie, 1998].

In U12-type introns, the roles of U1, U2, U4 and U6 snRNPs in U2-type introns are replaced by the U11, U12, U4atac and U6atac snRNPs, respectively [Wassarman and Steitz, 1992; Frilander and Steitz 1999; Patel and Steitz, 2003 ]. The overall similarity in the predicted secondary structure between analogous U2 and U12-type snRNAs suggests that the spliceosome rearrangements during catalysis are conserved between the two spliceosomes [Tarn and Steitz, 1997; Frilander and Steitz, 2001]. Moreover, the U12-type of introns is characterized by highly conserved consensus sequences at the donor and branch sites [Whu and Brendel, 2003]. Therefore, Burge *et al.*, 1998 designed a computer program, named *U12Scan*, to address the issue of the identification of U12-type introns. It is based on conserved motifs in the donor site and the branch site in the U12-type introns; they conducted a survey in a variety of species based on *GenBank* gene structure annotation. Later, Levine and Durbin [Levine and Durbin, 2001] adopted a slightly different strategy to recognize human U12-type introns. They first predicted U12-type introns in the human genome and then confirmed hypothetical introns using EST data *[how?]*. A *64bp* perfect match between a transcript sequence fragment and the *32bp* flanking sequences of predicted U12-type introns in both directions was required for confirmation. The latter approach does not suffer from the incompleteness or likely errors in the *GenBank* annotation, but has its own limitations. For example, any U12-type introns flanked by exons shorter than *32bp* would not be located. Both analyses restrict their search of DistBA *[undefined]*within a short region of the introns. This is consistent with the assumption that no U12-type introns have DistBAs shorter than *8nt* or longer than *21nt*.

The U2-type splicing signals have highly degenerate sequence motifs; many different sequences can function as U2-type splice sites. It is not clear how degenerate s[hozMequences at the splice sites of U2 intron types are recognized by spliceosomal complex. Clearly, if terminal dinucleotides were sufficient to be recognized, then GT-AG introns of U2 and U12-type introns would be excised by both U2 and U12 spliceosomes. however nothing of this kind has been described.

As it was defined before, the nature of protein–protein and RNA-protein cooperation is based on hydrophobic interactions [Kauzmann, 1959].

Matthew *et al.*, 2006 showed complex series of RNA-protein and RNA-RNA intimate associations at the heart of mammalian spliceosome during the course of splicing. *[HERE; this sentence does not mean anything: subject or main verb is missing]*Interestingly, examine RNA-protein and RNA-RNA interactions from point of view of assembly RNA and proteins with similar hydropathy values. Hydrophobic-hydrophilic properties of amino acids and nucleotides may be predicted on the base of half-empirical coefficients of hydropathy. [Guckian *et al.*, 2000]. We expect profiles based on hydropathy properties pave the road towards general understanding of recognition mechanisms of regions, which are essential for splicing, by spliceosome. Also we expect to use hydropathy profiles to distinguish U2 and U12-types of introns.

## METHODS AND PROGRAMS

In order to distinguish biochemically conservative regions around splice sites, we define a general hydropathy profile. Regions whose hydropathy differ from the background value are expected to be essential for recognition by spliceosome. The procedure is given below. We built a set of 313 introns of length greater than *200 nt* and a set of 385 exons of length greater than *60* *nt* from genes of 21$^{st}$ and 22$^{nd}$ chromosomes from <u>*GenBank* (http://www.ncbi.nlm.nih.gov).</u> Each gene contained 1 to 3 introns. We built a set <u>*T*</u> of flanking regions. The flanking sequences (*30 nt* within the exon and *30 nt* within the intron) were extracted at both exon–intron junctions, at 5'ss and 3'ss boundaries.

**Evaluating background hydropathy:** We determined the background frequency of the bases in exons and introns separately. We counted the bases distributions on 30 positions at the 5' and at the 3' boundaries of exons. There is no statistically confirmed difference between the two ones. Base frequencies were found to be *q(A) = 0.237, q(C) = 0.283, q(G) = 0.276* and *q(U) = 0.204*. For introns of lengths greater then 200, we counted the base distribution at the 5' boundaries only. Indeed, it is well known that a polypyrimidine track exists near the 3' boundaries of introns that do not allow to take them into account. Base frequencies were found to be *q(A) = 0.210, q(C) = 0.249, q(G) = 0.299* and *q(U) = 0.241*.

**Measure of hydropathy :** A hydropathy coefficient *hc(i)* may be associated to each base *i* . Coefficients provided in [Guckian et al., 2000], are *hc(A)=-1.07, hc(C)=-0.76, hc(G)=-1.36* and *hc(U)=-0.76*. Given a set of splice sites, one computes an average hydropathy value for each position as follows. For each base, its number of occurrences at a given position in the set is multiplied by its hydropathy coefficient. Summing over all the bases yields the average hydropathy value.

In order to exhibit a *specific profile*, we aligned all regions of our set <u>*T*</u> of flanking regions, that are *60 nt* long, at *5'ss* boundaries. An average hydropathy value was computed for each position of splice site set, using background frequencies and hydropathy coefficients given above. We proceeded similarly for regions at *3'ss* boundaries. This yields an average background value that is *-0.996* for exons and *-1.01* for introns. The corresponding variances are *0.0687* and *0.0690*. Given a position *i* on the *5'ss* or the *3'ss* boundary, the average hydropathy value can be interpreted as the sum of *k* random variables with a Bernoulli distribution given by the background frequencies. Therefore, its distribution is approximately normal. We denote *hp(0.99)* and *hm(0.99)* the two values that satisfy *P(E<h(i)<hp(0.99))=99%=P(hm(0.99)<h(i)<E)*. When *h(i)* is out of this range, we use the large deviation formula to compute the pvalue of *h(i)*.

$$log(P(h(i)>h) = -k(h-E)^2/2s.$$

We will compare two methods that identify common and distinguishing features in each splice-site type. We built four sets of 100 introns with two confirmed splice sites that were extracted from a user-friendly resource, <u>*SpliceRack*</u> (http://katahdin.cshl.edu:9331/SpliceRack/). Two sets are associated to human U2-type introns, with GT–AG termini for one set and GC–AG termini for the other. Two sets are associated to human U12-type introns with GT–AG termini for one set and AT–AC termini for the other. For each intron, one extracts a *38 nt* long region around each exon–intron junctions, at the 5'ss and at the 3'ss. More precisely, we extract *8 nt* within the exon and *30 nt* within the intron. These new sets are spread out on several chromosomes; therefore we use a different background. The background frequencies of nucleotides were evaluated from a set of 1228 human genes extracted from <u>*GenBank* (http://www.ncbi.nlm.nih.gov)</u> and were found to be *q(A) = 0.219, q(C) = 0.269, q(G) = 0.280* and *q(T) = 0.233*.

The flanking sequences of each set were separately aligned using _Pictogram_ program designed by Chris Burge and Frank White (http://genes.mit.edu/cgi-bin/pictogram.pl). _Pictogram_ is a handy tool to visualize sequence alignments and consensus sequences. Its input is an array of sequences of equal length. For each position _j_, it computes for each base _i_ its relative frequency, e.g. the ratio $p_i(j)/q_i$ of the frequency of each nucleotide $p_i(j)$ to the background frequency $q_i$. It also computes the information content that is defined as $\sum p_i(j) \log p_i(j)/q_i$. Information content is commonly used to study the variability (or level of conservation) at each position, varying from 2 (one event is certain) to 0 (all observed frequencies are equal to background probabilities) ([39]). _Pictogram_ output is a diagram of letters. For each position _j_, the height of _i_-th letter is proportional to the ratio $p_i(j)/q_i$. Additionally, _i_nformation content is written below. We evaluated the nucleotide conservation in the flanking regions of exons and introns with _Pictogram_ and our results are depicted in Figures 3 and 4.

We built hydropathy profiles for our four sets of splice sites in order to study and establish biochemical conservation of the flanking regions for each subgroup of U2 and U12-type introns with different dinucleotide boundaries.

## RESULTS

**Distinguishing biochemically conservative regions from background values**

Sliding along all regions of our set T, we compute an average hydropathy value for each position. Hydropathy value profile is illustrated in Figure 2. The termini of the introns are marked in red. The standard deviation depicted in vertical lines indicates the degree of conservation of biochemical properties for each nucleotide position of splice site.

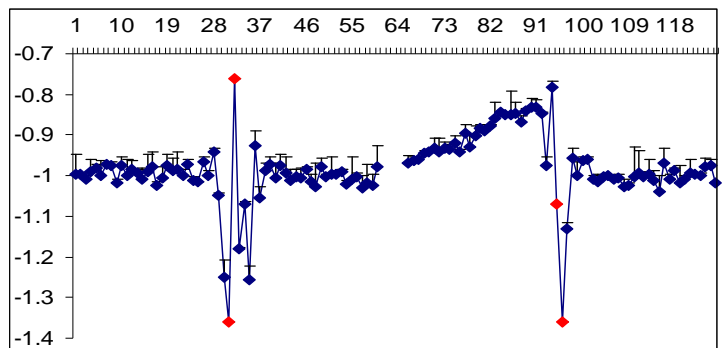

**Fig. 2. Distinguishing biochemically conservative regions from background values**
The numbers of nucleotides are marked on the _x_-axis and hydropathy values are indicated by the scale on _y_-axis.

Please, have figure and numbering consistent with explanations! Use number 0 above. Do NOT number from 1 to 37 but from -8 to 30 in Figures below. Explain where are the intronic and exonic parts. Then, I zill be able to rewrite the paragraph belowm which has no order...

Approximately normal base distribution corresponding the background hydropathy is observed at positions -30 to -3 within the introns and +8 to +30 within the exons at the 5'ss; at positions +2 to +30 within the exons at the 3'ss. *(MR: explain and give value).*

Regions at the positions -2 - +7 at the 5'ss and -30 - +1 at the 3'ss deviate from the background hydropathy. Slow decay at positions -30 to -5 due to pyrimidine abundance correspond to polypyrimidine track. This suggests that in these regions hydropathy values deviated from background hydropathy values. Therefore, they might correspond to binding sites for spliceosoma.

General hydropathy profile of all genes with 1- 3 introns from 21$^{st}$ and 22$^{nd}$ chromosomes resembles to the U2-type introns hydropathy profile which is shown above. This results from the low proportion of U12-type introns that does not exceed 0,34% [Levine and Durbin, 2001].

### Comparison of two methods of search potential binding sites of spliceosoma particles

In this section, we report results concerning a comparison of two methods that search potential binding sites of spliceosoma particles. The first method, now classical, is  the program *Pictogram* by Chris Burge and Frank Whit. It attempts to identify regions which have some nucleotide consensus. Our method attempts to point out regions which have conservative biochemical properties from a variable background. In Figures 3 and 4, the hydropathy profiles of U2 and U12- dependent introns with different termini and corresponding pictograms are depicted and may be compared. For all pictures of hydropathy profiles, the numbers of nucleotides are marked on the *x*-axis and hydropathy values are indicated by the scale on *y*-axis. The termini of the introns are marked in red. *The standard deviation of hydropathy at each position of splice sites indicates the degree of conservation of biochemical properties, namely hydropathy. It is depicted as vertical lines above the points that display the values of hydropathy at each position of splice site.* In pictograms the nucleotide consensus is displayed by the letter size that indicates the base frequency distribution at each position of the splice sites.

### U2-type splice sites:

100 splice sites with GT-AG and GC-AG termini extracted from *SpliceRack* were used to construct each pictogram. GT–AG and GC–AG subtypes of U2-type introns are considered separately in order to compare our results to the ones obtained by classical methods.

#### GT_AG U2-type introns

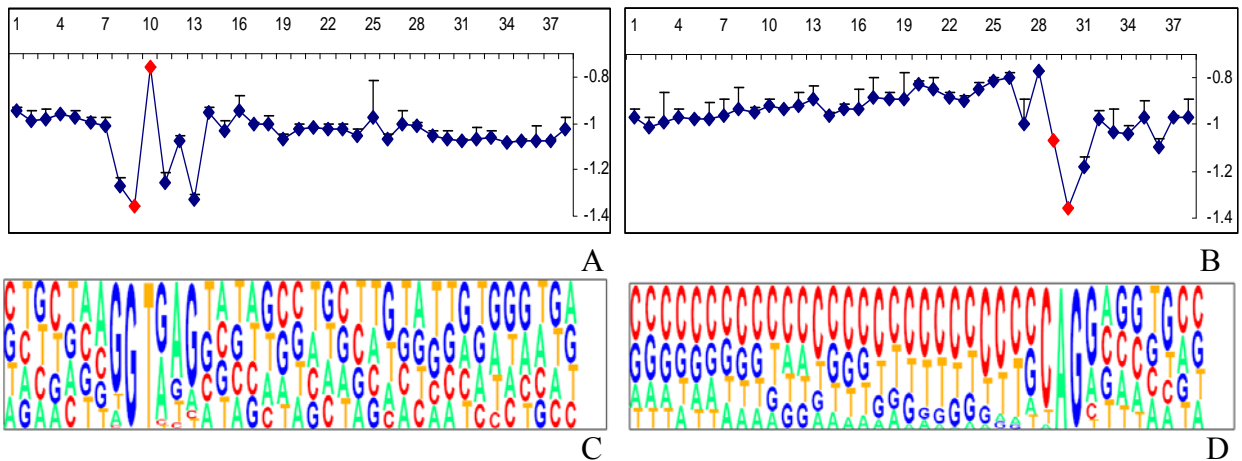

#### GC_AG U2-type introns

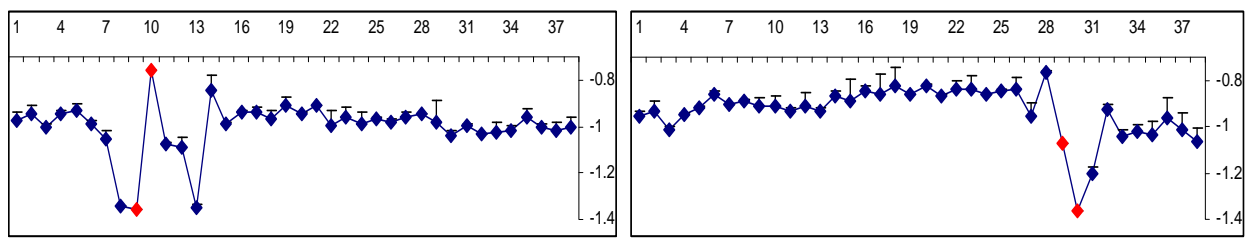

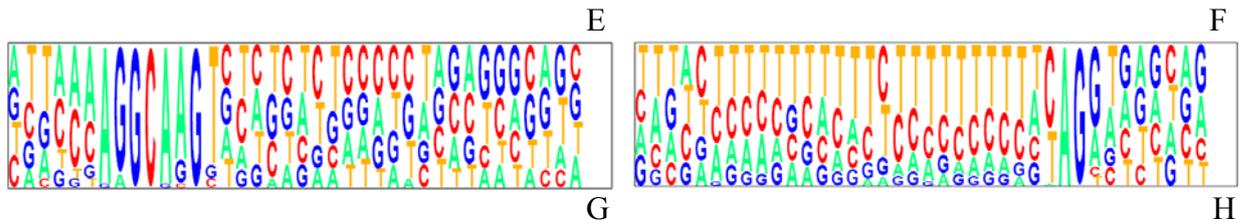

E                      F

G                      H

Figure 3. **The hydropathy profiles and corresponding pictograms for pairwise comparison of the U2-type introns.**
The hydropathy profiles and corresponding pictograms of the donor (right) and the acceptor (left) sites of two subtype s U2-type introns with GT-AG (**A-D**) and GC-AG (**E-H**) termini are shown for pairwise comparison.

Although the pictograms of GT–AG and GC–AG subtypes are significantly different (Fig. 3 C, D, G, H), the hydropathy profiles are quite similar. Indeed, nucleotide consensus of U2-type introns mainly contain quite hydrophilic purines when termini are either GT-AG (<span style="color:red">Figure 3C</span>) or and GC-AG (<span style="color:red">Figure 3C</span>). There is more conservation in the exonic nucleotides for 5'ss of U2-type GT–AG introns, compared to 5'ss of U2-type GC–AG introns. Moreover, the consensus sequence for 5'ss motif of U2-type GT–AG introns corresponds to a perfect base pairing to the U1 snRNA 5' end. In GC–AG introns, the substitution at position +2 of the 5'ss introduces a mismatch in the U1: 5'ss helix. The identity of hydropathy coefficients of T and C that are both hydrophilic pyrimidines apparently compensates the mismatch at position +2 of U2-type GC-AG introns. Nucleotides G and A can both be observed at the first intronic position of 5'ss of U2-type GT-AG introns. They are both competent for splicing as both are purines and have small values of hydropathy coefficients.

Therefore, our approach based on hydropathy evaluation provies an explanation for the competence of U2-type introns with high variability of nucleotides at the 5'ss for splicing machinery.

**U12-type splice sites:**

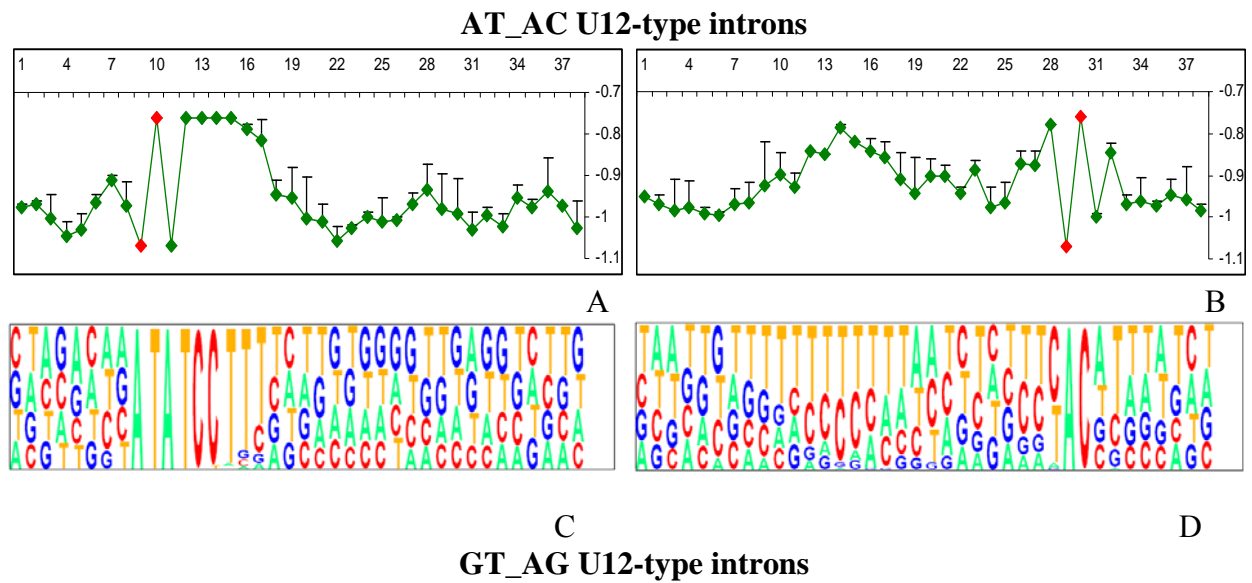

**AT_AC U12-type introns**

A                      B

C                      D

**GT_AG U12-type introns**

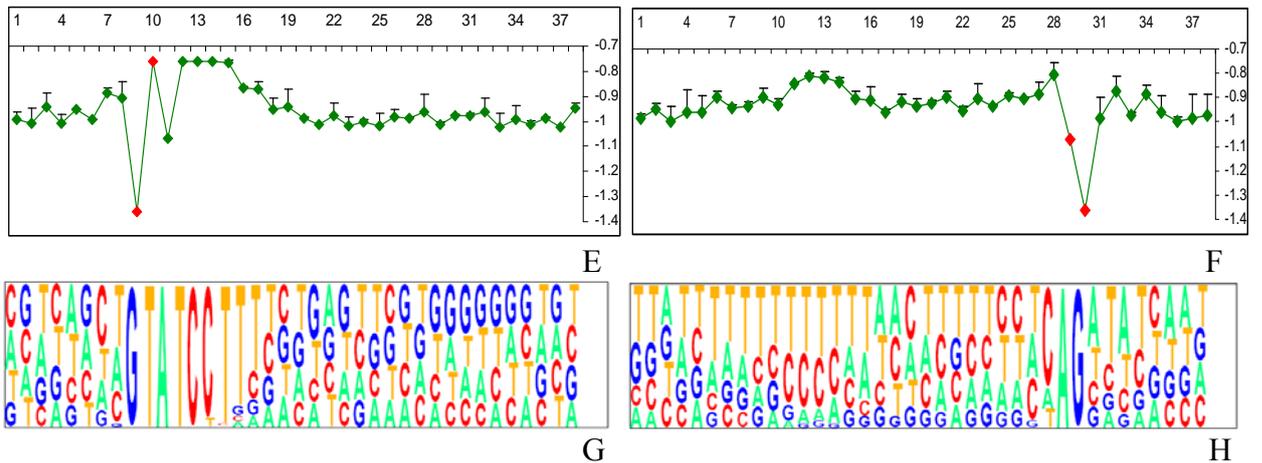

**Fig. 4. The hydropathy profiles and corresponding pictograms for pairwise comparison of the U12-type introns.**
The hydropathy profiles and corresponding pictograms of the donor (right) and acceptor (left) sites of two subtype s U12-type introns with AT-AC (**A-D**) and GT-AG (**E-H**) termini are shown for pairwise comparison.

In the pictograms of the 5'ss of U12-type introns (Fig. 3 C, G), the information content distribution for GT–AG and AT–AC 5'ss of U12-type introns shows a high degree of conservation at intronic positions that starts decaying at position +8. Indeed, this sequence consensus enriched by hydrophobic pyrimidines allows for a perfect distinction between splicing signal and its quite hydrophobic background starting at position +10. Additionally, a conservation of T at position –1 of the 5'ss of GT–AG U12-type introns can also be observed.

Pictograms method attempts to point out a nucleotide consensus and it can be observed from the figure 3D and H that U12-type introns have a very small degree of conservation in the intronic nucleotides of the 3'ss. Indeed the acceptor splice site of GT–AG is T- rich while GC–AG 3'ss is C- rich. Our method attempts to point out a consensus of biochemical properties, namely hydropathy. Indeed, the 3'ss of both subtype s of U12-type introns have resembling hydropathy profiles. Intronic nucleotides of both subtype s of U12-type introns are enriched in pyrimidines and as a result hydrophobic in spite of lack a PPT. This suggests these regions might be functionally very similar.

In order to explain recognition mechanisms of the U2 and the U12 – dependent introns with the same terminal dinucleotides we compared the hydropathy profiles of the U2 and the U12 – dependent introns (Fig. 2L and 3L). The 5'ss profiles for U12-type introns are different from the U2-type introns ones. As pointed out above hydropathy profiles of the 5'ss of U12- dependent introns have hydrophobic canonical pyrimidine rich ATCCTTT motif (plateau in Figure 3 A and E) that succeeds terminal dinucleotides, but exonic and intronic nucleotides of the 5'ss of U2 – dependent introns are purine rich and hydrophilic as consequence. We believe it explains the fact that U2 and U12 – dependent introns with same terminal dinucleotides are excised by different type s of spliceosomas.

## Conclusion

We expected a high degree of biochemical properties conservation due to the existence of consensus sequences at the 5'ss and the 3'ss. Our graphics demonstrate that exist quite long biochemically conservative regions at the splice sites of U2 as well despite the lack of consensus for sequences.
We show that there is a similarity of hydropathy profiles inside intron types. On the one hand, GT–AG and GC–AG introns belonging to U2-type have resembling hydropathy profiles as well as AT–AC and GT–AG introns belonging to U12-type. On the other hand, hydropathy profiles of U2 and U12-types GT–AG introns are completely different. Our analysis should be a step forward for a general understanding of recognition of regions, which are essential for splicing, by spliceosome and for a distinction of U2 and U12-types of introns.